\documentclass[preprint,12pt,longtitle,authoryear]{elsarticle}

\usepackage[margin=1.0in]{geometry} 
\usepackage{multicol}

\usepackage{epsfig}
\usepackage{latexsym}
\usepackage{epic}
\usepackage[leqno]{amsmath}
\usepackage{amssymb}

\biboptions{numbers}

\usepackage{fullpage}
\usepackage{amsmath}
\usepackage{amsfonts}
\usepackage{amssymb}
\usepackage{latexsym}
\usepackage{ifthen}
\usepackage{theorem}
\usepackage{graphicx}
\usepackage{array}
\usepackage{url}
\usepackage{setspace}
\doublespace
\newtheorem{prop}{Proposition}[section]
\newtheorem{lemma}[prop]{Lemma}
\newtheorem{theorem}[prop]{Theorem}

\newcommand{\EE}{\mathbb{E}}
\newcommand{\PP}{\mathbb{P}}

\begin{document}
\begin{frontmatter}

\title{The impact and interplay of long and short branches on phylogenetic information content}
\author{Iain Martyn and Mike Steel$^*$}
\ead{iain.martyn@mail.mcgill.ca, mike.steel@canterbury.ac.nz}
\address{Biomathematics Research Centre, \\University of Canterbury, Christchurch, New Zealand}



\date{\today}

\begin{abstract}
In molecular systematics, evolutionary trees are reconstructed from sequences at the tips under simple models of site  substitution.  A central question is how much sequence data is required to reconstruct a tree accurately? The answer depends on the lengths of the branches (edges) of the tree, with very short and very long edges requiring long sequences for accurate tree inference, particularly when these branch lengths 
are arranged in certain ways.  For four-taxon
trees, the sequence length question has been investigated for the case of a rapid speciation event in the distant past.  Here, we generalize results from this earlier study, and show that the same sequence length requirement holds even when the speciation event is recent, provided that at least one of the four taxa is distantly related to the others.  However, this equivalence disappears if  a molecular clock applies, since the length of the long outgroup edge becomes largely irrelevant in 
the estimation of the tree topology for a recent divergence.
  We also discuss briefly some extensions  of these results to models in which substitution rates vary across sites, and to settings where more than four taxa are involved.  
  \end{abstract}

\end{frontmatter}


$^*$ Corresponding author

\newpage

\section{Background}

Phylogenetic methods are founded on the notion that evolutionary relationships can be inferred from sequences that have evolved along with the taxa. It is usually supposed that such sequences evolve according to some continuous-time reversible Markov process, or a mixture of such processes (for further background on phylogenetic inference, the reader is referred to  \cite{Fels}). Here, we are interested in the question of  the sequence length required to accurately estimate a discrete and fundamental parameter of evolutionary history, namely  the topology of the underlying evolutionary tree.  This question has long been of interest in molecular systematics (see, for example, \cite{sai}, 
\cite{chu} and \cite{lec}) and a variety of mathematical approaches have been explored in order to quantify how much `phylogenetic information' sequence data contains  (\cite{mos},  \cite{tow07},  \cite{tow11} and \cite{tow12}).  Although the underlying tree topology is rooted, phylogenetic models are generally time-reversible, and so methods based on these models produce trees that are unrooted; accordingly, we will say that a
method correctly reconstructs the tree topology if it does so up to the placement of the root. 

Amongst unrooted trees,  the simplest phylogenetic problem involves a set of four 
 taxa, for which there are just three resolved binary tree topologies and one `star tree'.   \cite{FS} investigated the sequence length required to accurately reconstruct a  binary four-taxon phylogenetic tree with four long pendant branches, and a short interior edge  (Fig. 1(a), 1(a$'$)).  This special case is motivated by the scenario in evolutionary biology in which a rapid speciation event in the distant past results in all taxa sitting on `long branches' around a short interior edge of length $l_0$.   The authors found that the length of  sequence needed to reconstruct the correct four-taxon tree with probability $1- \epsilon$ grows at the rate $C e^{bL}/l_{0}^2$, where $C$ and $b$ are positive constants and  $L$ is the length of the long pendant edges.   Notice the impact on the required sequence length of a long branch length (i.e. $e^{bL} \rightarrow \infty$ as $L \rightarrow \infty$) and of a short interior branch (i.e. $\frac{1}{l_0^2} \rightarrow \infty$ as $l_0 \rightarrow 0$), and that these combine multiplicatively in this lower bound (thus, the cumulative effect of a short branch beside a long one becomes compounded much more than if the interaction was, say, additive).   This formally justifies the informal notion that a very short interior edge surrounded by long branches is a particularly challenging phylogenetic problem.

In this paper, we wish to compare this scenario with another that is at least as common in evolutionary biology, namely the setting in which only one of the taxa is distantly related to the others, being a distant `outgroup' taxon  (see, for example, Fig. 1(b)).    In particular, we ask whether the sequence length requirements are less severe if just one pendant edge is long, rather than all four.
We show analytically that essentially the same bound (of the form $Ce^{bL}/l_{0}^2$) applies in general.  

However, a curious situation develops if one imposes a molecular clock.  Doing so does not affect the exponential depending on $L$ of the sequence length requirements for reconstructing the tree in Fig. 1(a$'$).  However, in the case of just one distant outgroup taxon (Fig. 1(b$'$)), the exponential dependence on $L$ (the long branch) disappears entirely.  

Finally, we extend our main  result to settings where sites evolve at varying rates and we also indicate how the results apply when the four lineages are replaced by four monophyletic groups of taxa.

 \begin{figure}[h] \begin{center}
\resizebox{12cm}{!}{
{
\includegraphics{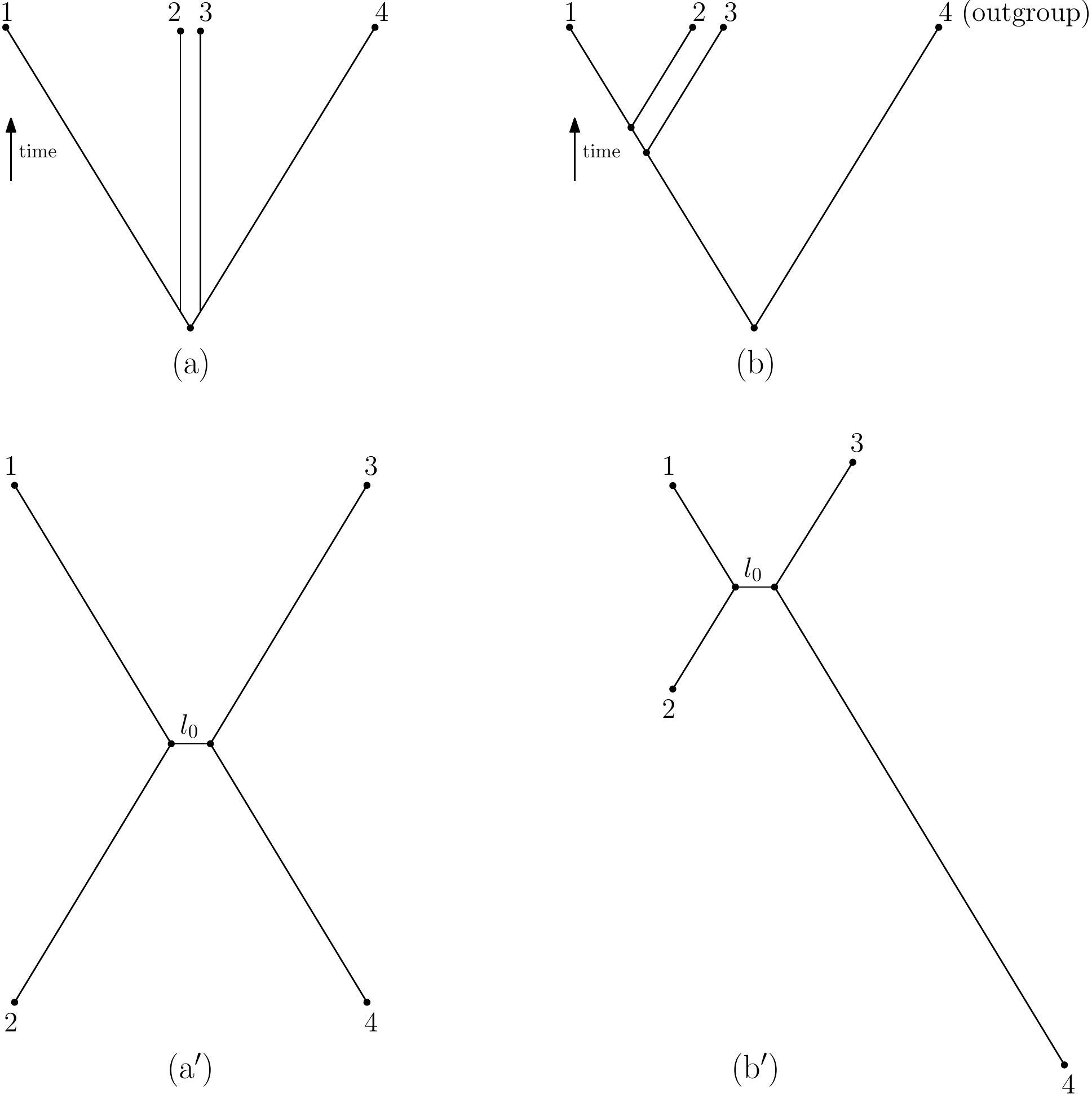}}
}
\caption{In tree (a), a short interior edge is incident with four long pendant edges, representing a rapid radiation event deep in the past; tree (a$'$) shows the associated unrooted tree.  Tree (b) shows a more recent rapid radiation event in which only one of the four incident edges is  long, as it joins a distant outgroup. Tree (b$'$) shows the associated unrooted tree.}
\end{center}
\label{fig1}
\end{figure}

\section{Preliminaries}
We first recall some terminology from phylogenetics.  Let $X=\{1,2,3,4\}$ be a set of four taxa, and let $ T _{1},T _{2}$ and $T _{3}$ be the three possible unrooted binary trees that have $X$ as their leaf set.  

Suppose we have a continuous, stationary, and time-reversible  Markov process on a state space $G$ that acts at various intensities on the edge of one of these trees.
 The {\em length} of an edge will refer to the expected number of substitutions on that edge.  This is the substitution rate on that edge, multiplied by the 
temporal duration of that edge.  In the case where the substitution rate is constant across the tree, we will say that a {\em molecular clock} applies, but we do not assume this unless otherwise stated. 
Throughout this paper we will let  $l_{0}$ be the branch length of the interior edge of any four-taxon tree.

Let $S=G^X$ be the set of of possible assignments of elements of the state space $G$ to the leaf set $X$; we will refer to an element of $S$ as a {\em site pattern}.
Now,  suppose we generate $k$ site patterns independently according to the same Markov process  to form sequences of length $k$ (one sequence for
each taxon).  It is well known that for any set of (positive) branch lengths on $T_i$, one can correctly recover the topology $T_i$ from these sequences with a probability of at least $1-\epsilon$ for sufficiently large values of  $k$  (for a discussion of statistically consistent tree reconstruction, see \cite{Fels}). 
Here `sufficiently large' depends not just on $\epsilon$ but also on the tree and its associated branch lengths.

As in \cite{FS}, our arguments rely on the properties of  the {\em Hellinger distance}  ($d_H$),  which is defined  as follows:
Given a finite set $U$, the Hellinger distance $d_H(p,q)$ between two probability distributions $p$ and $q$ on $U$ is defined by the equation:
\begin{equation}
d^2_{H}(p,q)=\sum_{u\in U} (\sqrt{p_{u}}-\sqrt{q_{u}})^2 = 2(1- \sum_{u\in U} \sqrt{p_uq_u}).
\label{equation:Hell1}
\end{equation}

Hellinger distances are useful to quantify the amount of data required to accurately identify a discrete parameter in a stochastic model.  We will describe this below (Lemma~\ref{lemma:SZE}) in a general setting, not specific to phylogenetics.

\subsection{Hellinger distance bounds on required sequence length}

Let $A$  and $U$ be finite sets,  and suppose that each element $a \in A$ defines a probability distribution $p_a$ on $U$.
We will denote the Hellinger distance between $p_a$ and $p_b$ by $d_H(a,b)$, and, by slight abuse of terminology, refer to it as the `Hellinger distance between $a$ and $b$'.  

Suppose an element $\xi$ of $A$ is selected according to some discrete non-zero probability distribution on $A$.    Conditional on $\xi=a$, consider a sequence of $k$ samples of $U$ generated independently in $U$ according to the probability distribution $p_a$. 
Let $M: U^k \rightarrow A$ be some 
method for estimating the element $a \in A$ from a sequence $(u_1, u_2, \ldots, u_k) \in U^k$ (here $M$ may be a deterministic function from $U^k$ to $A$ or a process that selects an element of $A$ from each element of $U^k$ according to some probability distribution -- the latter case allows ties to be broken randomly).

Let $r^{(M, k)}_a$ denote the probability that the method $M$ correctly identifies the element $a$ 
that generates the sequence $(u_1, u_2, \ldots, u_k) \in U^k$ under the probability distribution $p_a$.   In other words:
$$r_a^{(M, k)}= \PP(M(u_1, u_2, \ldots, u_k) = a| \xi =a).$$

The following lemma is from \cite{SZE} (Theorem 3.1 and (2.7)).

\begin{lemma}
\label{lemma:SZE}
Given finite sets $U$ and $A$, suppose that 
elements of $U$ are generated i.i.d. by some unknown element $\xi \in A$.  Then for any estimation method $M$ that satisfies $r^{(M, k)}_{a}\geq1-\epsilon$, for all $a \in A$, 
we must have  $$k\geq \frac{C_{\epsilon}}{d^2},$$
where $C_{\epsilon}= \frac{1}{4}(1-\frac{|A|}{|A|-1}\epsilon)^2$,
and 
$d = \min \{d_H(a, a'):  a,a' \in A; a \neq a' \}.$

\end{lemma}

In our setting, $A$ will consist of a set of phylogenetic trees on the leaf set $X=\{1,2,3,4\}$  and $U$ will be the set $S$ of assignment of states of the elements of $X$.      We will use  the lemma to prove a lower bound on $k$ in the following Section.

\section{A general lower bound on the required sequence length}
\label{sec:General Markov Process}

We now present a lower bound for the necessary sequence length $k$ required to reconstruct a tree of the type shown in Fig. 1(b). 
This lower bound is essentially of the same form as that which applies when all four pendant branches are long -- namely, it grows exponentially with the length $L$ of the long branch and in inverse proportion to the square of the short interior branch, and these factors combine multiplicatively.

\begin{theorem}
\label{thm1}
Consider the three-leaf star tree on the taxon set $\{1,2,3\}$ with corresponding branch lengths $l_{1},l_{2},l_{3}\geq \delta>0$.  Suppose that a fourth taxon is attached by a branch of length  $L>0$ to one of the three branches at a distance $l_0\in (0, \delta)$ from the interior node.   Generate $k$ i.i.d. site patterns at the tips of the resulting four-taxon tree under a finite-state, stationary and irreducible Markov process.  Then, any method that is able to correctly identify with probability at least $1-\epsilon$ which branch the fourth branch is grafted onto requires:
\begin{equation}
\label{lowerB}
k \geq  C e^{bL}/l_{0}^2
\end{equation}
where $C$ is a constant (independent of $l_0$ and $L$) that depends on $\epsilon, \delta$ and  the rate parameters of the Markov process.

Moreover, some methods achieve this accuracy using sequences with a length that is no more than a constant times  $e^{bL}/l_{0}^2$.

\end{theorem}

 \begin{figure}[h] \begin{center}
\resizebox{12cm}{!}{
{
\includegraphics{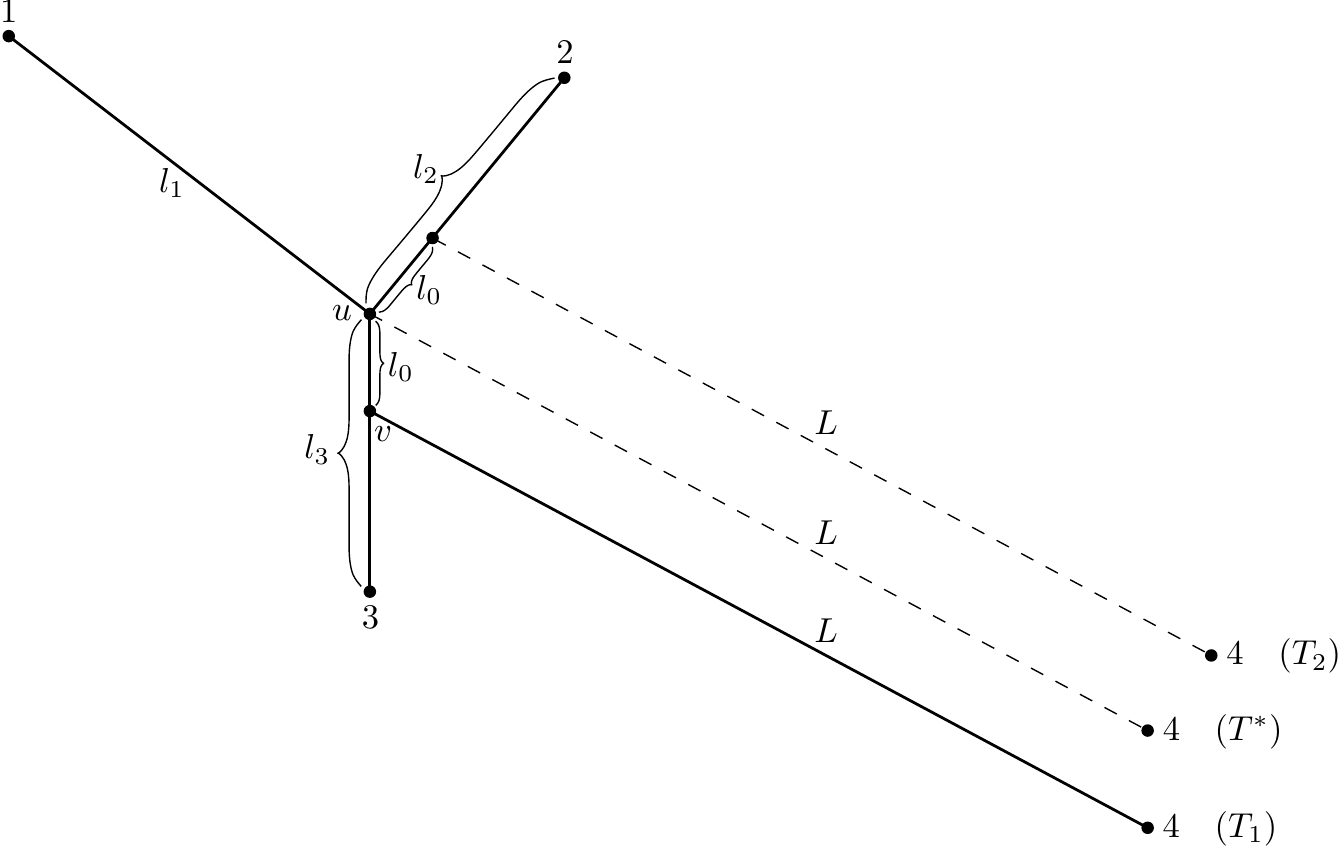}}
}
\caption{The three-taxon and four-taxon trees described in the statement and proof of Theorem~\ref{thm1}}
\end{center}
\label{fig2}
\end{figure}

{\em Proof: }  To establish Inequality (\ref{lowerB}) as a lower bound,  we first derive an upper bound for the Hellinger distance between $T_1$ and the tree $T^*$, formed by grafting the fourth branch directly onto vertex $u$ of the three-taxon star tree, as shown in Fig.~2.  By the triangle inequality, we have:

\begin{equation}
d_{H}(T_{1},T_{2})\leq  d_{H}(T_{1},T^*) +d_{H}(T^*,T_{2}).
\end{equation}
Most of the proof is devoted to establishing the following inequality (for a constant $B = B(\delta)$):

\begin{equation}
d^2_{H}(T_{1},T^*)\leq B l_0^2 e^{-bL}.
\label{equation:eq2}
\end{equation}

To establish Inequality (\ref{equation:eq2}), let $p_s$ and $p_s^*$ denote the probability of generating the site pattern $s$ on $T_1$ and $T^*$, respectively.
Let  $N$ denote the number of substitutions occurring along the edge between $u$ and $v$  (the interior edge of $T_1$), and let  $\tau=\mathbb{P}(N>0)$.
Further, let $Q_s$ (respectively $Q^*_s$) denote the conditional probability of generating pattern $s$ on $T$ (respectively, on $T^*$) given that $N=0$.
Similarly,  let $P_{s}$ (respectively $P_s^*$) denote the conditional probability of generating pattern $s$ on $T$ (respectively, on $T^*$)  given $N>0$.  Then, by the law of total  probability, we can write $p_s$ and $p^*_s$ as:

\begin{eqnarray}
p_s &=& (1-\tau)\cdot Q_{s}+\tau\cdot P_{s}, \\
\nonumber
p_s^* &=& (1-\tau)\cdot Q_{s}^*+\tau\cdot P_{s}^*. \\
\nonumber
\end{eqnarray}

Let $D_s=P_s- P_s^*$.  Then, since $Q_s=Q_s^*$, we have:

\begin{equation}
\label{hel}
p_s-p_s^*=\tau(P_s- P_s^*)=\tau D_s.
\end{equation}

Now, the Hellinger distance between $T_1$ and $T^{*}$ (on site patterns) is:
\begin{equation}
d^2_{H}(T_{1},T^*)=2(1-\sum_{s\in S} \sqrt{p_{s}p^*_{s}}).
\label{equation:Hell}
\end{equation}
Following the approach of \cite{FS} (Lemma 5.1),  substituting $p^*_s = p_s - \tau D_s$ (from Eqn. (\ref{hel})) into (\ref{equation:Hell}) gives:
$$d^2_{H}(T_{1},T^*)=2\left(1-  \sum_{s\in S}p_s\sqrt{1 - \frac{\tau D_s}{p_s}}\right) $$
and then  the application of the inequality $\sqrt{1+y}\geq 1+ y/2-{y^2}/{2}$ for $y \geq -1$   leads to the following inequality:
\begin{equation}
\label{intermedeq}
d^2_{H}(T_{1},T^*)\leq  \tau^2\cdot \sum_{s\in S}\frac{D_s^2}{p_s}.
\end{equation}
Now,  $\tau=\mathbb{P}(N>0)\leq\mathbb{E}(N)$, so we have  $\tau\leq l_0$, and thus we can replace $\tau$ in (\ref{intermedeq}) by $l_0$ to obtain:
\begin{equation}
d^2_{H}(T_{1},T^*) \leq l_0^2\cdot \sum_{s\in S}\frac{D_s^2}{p_s}.\\
\label{equation:eq1}
\end{equation}

Referring again to Fig. 2, let $\chi_u$ be the character state on vertex $u$, let $\chi_v$ be the character state on vertex $v$, and let $\chi_1, \ldots, \chi_4$ be the respective character states on the leaf set \{1,2,3,4\}, and let $p(s,s')$ denote the conditional probability $\mathbb{P}(\chi_u=s',\chi_v=s''|N>0)$.

Then we have:
$$P_s = \sum_{{s', s''\in S}}p(s,s')\mathbb{P}(\chi_1=s^1|s')\mathbb{P}(\chi_2=s^2|s')\mathbb{P}(\chi_3=s^3|s'')\mathbb{P}(\chi_4=s^4|s''),$$
and
$$P_s^* = \sum_{{s', s''\in S}}p(s,s')\mathbb{P}(\chi_1=s^1|s')\mathbb{P}(\chi_2=s^2|s')\mathbb{P}(\chi_3=s^3|s'')\mathbb{P}(\chi_4=s^4|s'),$$
where 
$ \mathbb{P}(\chi_i=s^i|x)$ (for $x= s'$ or $s''$) is probability of generating leaf state $s_i$ at leaf $i$ conditional on state $x$ at the vertex of the tree adjacent to leaf $i$.

Thus,  $|D_s|= |P_s -P_x^*|$ is  bounded above as follows:
\begin{eqnarray}
\label{eqnarray:diff}
\mbox{    } |D_s| \leq  \sum_{{s', s''\in S}}p(s,s')\mathbb{P}(\chi_1=s^1|s')\mathbb{P}(\chi_2=s^2|s')\mathbb{P}(\chi_3=s^3|s'')\cdot \\
\newline
\bigg|\bigg(\mathbb{P}(\chi_4=s^4|s'')-\mathbb{P}(\chi_4=s^4|s')\bigg)\bigg|
\nonumber
\end{eqnarray}

We now invoke the property that any irreducible Markov process converges to its stationary distribution at an exponential rate regardless of its starting state ({\em cf. } Theorem 8.3 of  \cite{Roz}).  Specifically, if $Y_t$ is the state of such a process when it is run for duration $t$ then, for any state $s$ with the equilibrium frequency $\pi(s)$, and any second state $\alpha$, we have: 
\begin{equation}
|\mathbb{P}(Y_t=s|Y_0=\alpha)-\pi(s))|\leq Ae^{-a t},
\label{eqn:ineq1}
\end{equation}
where $A$ and $a$ depend only on the rate parameters of the Markov process.  Using the triangle inequality, Inequality  (\ref{eqn:ineq1}) gives:
\begin{equation}
|\mathbb{P}(\chi_4=s^4|s')-\mathbb{P}(\chi_4=s^4|s'')| \leq 2Ae^{-aL}.
\label{equation:pullout}
\end{equation}

Substituting Eqn. (\ref{equation:pullout}) into Eqn. (\ref{eqnarray:diff}), we obtain:
\begin{equation}
\label{fraceq}
\frac{D_s^2}{p_s}\leq \frac{4A^2e^{-2aL}}{p_s}\bigg( \sum_{{s', s''\in S }}p(s,s')\mathbb{P}(\chi_1=s^1|s')\mathbb{P}(\chi_2=s^2|s')\mathbb{P}(\chi_3=s^3|s'')\bigg)^2
\end{equation}
and since the term in brackets  is bounded above by  $1 (= \sum_{{s', s''\in S }}p(s,s'))$, we obtain:
\begin{equation}
\label{Dineq}
\frac{D_s^2}{p_s}\leq \frac{4A^2e^{-2aL}}{p_s}.
\end{equation}
Note also that, since   $l_{1}, l_{2}, l_{3}\geq \delta>0$ and $L>0$, and the Markov process is irreducible, there is some positive $\rho = \rho(\delta)$ such that  $p_s\geq \rho$.  We can thus further reduce Inequality (\ref{fraceq}) to:
\begin{equation}
\frac{D_s^2}{p_s}\leq \frac{4A^2e^{-2aL}}{\rho}.
\label{equation:withd}
\end{equation}
Substituting Eqn. (\ref{equation:withd}) into Eqn. (\ref{equation:eq1}) now furnishes the promised justification of  Eqn. (\ref{equation:eq2}), upon taking
$b=2a$ and $B=B(\delta) = 4A^2/\rho(\delta)$.  By symmetry,  Eqn. (\ref{equation:eq2}) gives us the same upper bound on $d^2_{H}(T_{2},T^*)$ as for $d^2_{H}(T_{1},T^*)$.
We then have, by the triangle inequality:
\begin{equation}
\label{helps3}
d_{H}^2(T_{1},T_{2})\leq  (d_{H}(T_{1},T^*) +d_{H}(T^*,T_{2}))^2< 4Bl_0^2e^{-bL}.
\end{equation}
The first part of Theorem~\ref{thm1}  follows from  Lemma \ref{lemma:SZE} by taking $A = \{T_1, T_2\}$ (so that   $C_\epsilon= \frac{1}{4}(1-2\epsilon)^2$) and then setting
$C = \frac{C_{\epsilon}}{4B(\delta)}$. 

Finally, the last claim in Theorem~\ref{thm1} (that  $e^{bL}/l_{0}^2$ is an upper bound on the required sequence length, up to a constant multiplicative factor)  is provided by Theorem 14 of  \cite{logs}.

\hfill$\Box$

\subsection{Imposing a (relaxed) molecular clock}
\label{sec:3branches}
When a molecular clock is imposed,  there is an interesting shift in the sequence length requirements for accurate tree reconstruction.  Although we have seen that the two scenarios in Fig. 1 lead to the same type of lower-bound dependence of sequence length on $l_0$ and $L$, namely $\exp(cL)/l_0^2$, if we impose a molecular clock,
then this equivalence disappears.  More precisely, it is clear (from \cite{FS}) that the term $\exp(cL)/l_0^2$ remains for the deep divergence set-up of our Fig.1(a), but for the recent divergence event shown in Fig. 1(b) we will show that the length of the long edge $L$ is largely irrelevant.  

We need to stress here how this result should be interpreted. We are not claiming that if a clock applies in the tree that generates the data, then every consistent model-based method, such as maximum likelihood, will be immune to the effect of a long branch to an outgroup.  It will not be so immune if, in the model assumed in the maximum likelihood analysis,  a molecular clock is not imposed. We are merely claiming that certain methods (such as agglomerative clustering, or MLE with a clock) can be immune to a long branch if a clock assumption applies.

We formalize this by a result, in which the full strength of the molecular clock condition can be relaxed slightly. Note that for the tree in Figure 1b, then under a strict molecular clock
the branch lengths (as indicated in Fig. 2) must satisfy: $$l_1=l_2; \mbox{ } l_3=l_1+l_0 \leq L.$$ We relax this slightly by requiring only that:
\begin{equation}
\label{molclockeq}
\min\{l_3, L\} \geq  \max\{l_1, l_2\} +l_0.
\end{equation}

\begin{theorem}
\label{clocks}
Consider the tree in Fig. 1(b) and suppose that the branch lengths (as indicated in Fig. 2) satisfy the relaxed clock condition described in (\ref{molclockeq}).
Let  $k$ sites evolve i.i.d along this tree under a finite-state, stationary and reversible Markov process.  Then the placement of the branch leading to taxon 4 can be determined  correctly with probability at least $1-\epsilon$ provided that:
\begin{equation}
\label{Beq}
k \geq B/(1-e^{-\lambda l_0})^2,
\end{equation}
 where $B$ depends just on  $l_1$, the model and $\epsilon$,  and where $\lambda$ is a constant determined by the model. 
In particular, this bound is independent of the length $L$ of the long branch to the outgroup taxon $4$.

\end{theorem}
{\em Proof: }
Consider the following simple reconstruction method (this is, essentially,  unrooted UPGMA on four taxa). Let $s(x,y)$ denote the proportion of sequence sites for which taxa $x$ and $y$ have the same state. Select the two taxa that maximize $s(x,y)$ and return the (unrooted) quartet tree in which $x$ and $y$ form a cherry. 
Let $e(x,y)$ be the expected value of $s(x,y)$. If $Y_t$ $(t \geq 0)$, denotes the Markov process described in the statement of Theorem~\ref{clocks} then we have:
$$e(x,y) = \EE[s(x,y)]= \sum_i \pi_i \PP(Y_t=i|Y_0=i),$$
where in this equation that value $t= t_{xy}$ refers to the branch length distance between taxon $x$ and $y$.
By the spectral representation of reversible continous-time Markov processes (see e.g. Chapter 3, Eqn (40) of \cite{Ald})  
we have, for any state $i$:
$$\PP(Y_t=i|Y_0=i) = \pi_i + \sum_{m \geq 2} u_{im}^2 e^{-\lambda_m t},$$
where $\pi_i$ is the equilibrium frequency of state $i$, $\lambda_m\geq 0$ are the eigenvalues of the rate matrix multiplied by -1, and the $u_{jm}$ values are real coefficients related to the eigenvalues of the rate matrix.   The $\lambda_m$ values can be ordered $0=\lambda_1< \lambda_2 \leq \lambda_3\leq \cdots$.
Consequently, 
$e(x,y) = \sum_i \pi_i^2 + \sum_{m \geq 2} c_m e^{-\lambda_m t},$
where $c_m = \sum_i \pi_i u_{im}^2 >0,$
and so:
$$e(x,y) - e(x', y') = \sum_{m \geq 2} c_m (e^{-\lambda_m t_{xy}} - e^{-\lambda_m t_{x'y'}}).$$
Since the coefficient $c_2$ is strictly positive, if $t_{x'y'}-t_{xy}  \geq l_0$ we can write:
\begin{equation}
\label{Eeq}
e(x,y) - e(x',y')  \geq c_2 e^{-\lambda t_{xy}}(1-e^{- \lambda l_0})>0,
\end{equation}
where, for convenience, we let $\lambda$  denote $\lambda_2$.
Notice that $t_{12}=l_1+l_2$,and $t_{13}=l_1+l_3$ and $t_{23}= l_2+l_3$ and so, by the relaxed clock condition (\ref{molclockeq}) we have
$t_{23}-t_{12} \geq l_0$ and $t_{13}-t_{12} \geq l_0$. Thus (\ref{Eeq}) holds for $(x,y) = (1,2)$ and $(x',y') = (1,3), (2,3)$.

Next, if we $X_{12;3} =s(1,2)-s(1,3)$, then observe that:
\begin{equation}
\label{helpsplenty}
 \PP(s(1,2)< s(1,3)) = \PP(X_{12;3}<0)=\PP(X_{12;3} - \EE[X_{12;3}] < - \EE[X_{12;3}]).
 \end{equation}
 In order to exhibit an upper bound this probability, we will apply McDiarmid's inequality \citep{mcd}.  First, observe that we can express $s(1,2)-s(1,3)$ as a sum of $k$ independent random variables (one for each site), each taking a value of $+1, 0$ or $-1$, and this sum has the property that changing any one of these variables (while keeping the others fixed) alters $s(1,2)-s(1,3)$ by  an additive factor whose absolute value is at most $2/k$.  
Applying the McDiarmid inequality, noting that: 
$e(1,2)-e(1,3)\geq  c_2e^{-\lambda t_{12}}(1-e^{-\lambda l_0})$
from Inequality (\ref{Eeq}), we obtain, from Eqn. (\ref{helpsplenty}) that:
$$\PP(s(1,2)< s(1,3))  \leq \exp(-k c_2^2 e^{-2\lambda t_{12}}(1-e^{-\lambda l_0})^2/2),$$
and this can be made less or equal to $\epsilon/5$ whenever  Inequality (\ref{Beq}) is
satisfied for $B = \frac{2 \ln(5/\epsilon)}{c{_2}^2e^{-2\lambda t_{12}}}.$
By symmetry,
$\PP(s(1,2)< s(2,3)) $ is also less or equal to $\epsilon/5$ for this value of $k$. Moreover, by  the relaxed clock condition, we also have:
$$\PP(s(1,2) < s(x, 4)) \leq  \epsilon/5 \mbox{ for } x=1,2,3.$$

Thus, with probability at least $1-\epsilon$, the pair $\{1,2\}$ will have the strictly largest $s$-value;  consequently,  the correct tree topology will be recovered by the method described with probability at least $1-\epsilon$.

\hfill$\Box$

\section{Further extensions and concluding comments}

\subsection{Rates across sites}

When sites evolve i.i.d. the sequence length required to reconstruct the tree in Fig. 1(a) accurately grows  exponentially with the length of $L$ of the long exterior branches; the same holds also for the tree in Fig. 1(b) in the absence of any molecular-clock assumption (Theorem~\ref{thm1}).  We point out that these conclusions need not hold when the sites evolve independently but not identically under a model that allows substitution rates to vary across sites, provided this rate distribution allows arbitrarily small rates, and with appropriate density. 
Suppose, for example, that site $i$ has rate $r_i = \frac{1}{i} \mbox{ for }  i = 1, 2, \ldots.$
Let  $T'$ be either an alternative binary tree to $T_1$ or the unresolved tree (i.e. $T' = T_2$ or $T^*$), and let $D_H^2(T_1, T')$ be the Hellinger distance
between {\em sequences} of length $k$ generated by $T_1$ and $T'$ in which the rates at site $i$ of the Markov process is $r_i$. 
We claim that  for a sequence length that grows at the (polynomial) rate $L^5$,   the value $D_H^2(T_1, T')$ converges to $2$ as $L$ tends to infinity. 
We first establish this claim and then explain why it implies that one can reconstruct the generating tree ($T_1$) from sequences of a length that is polynomial in $L$.

By a standard equality relating Hellinger distance of sequences of independent samples to the Hellinger distances at each sequence site (easily derived from Eqn. 
(\ref{equation:Hell1}))  we have:
\begin{equation} 
\label{DHeq}
D_H^2(T_1, T') =2\left(1-\prod_{i=1}^{k} (1-\frac{1}{2}d_i^2)\right),
\end{equation}
where $d_i$ is the Hellinger distance between the probability distributions on patterns at site $i$ (and rate $r_i$) generated by tree $T_1$ and generated by $T'$. 
This applies in either the setting of Figure 1a (four long pendant edges) or Figure 1b (one long  pendant edge).   
Moreover, by definition,
\begin{equation}
d_i^2 \geq (\sqrt{p_i} - \sqrt{p'_i})^2
\label{exq2}
\end{equation}
where $p_i$ here refers to the probability of generating a site pattern with leaves $1,2$ in one state (say $A$) and leaves $3,4$ in a different state (say $B$) 
on $T_1$ at substitution rate $\lambda_i$,  while
$p'_i$ is the corresponding probability for this same site pattern when $T_1$ is replaced by $T'$.  Notice that this site pattern can be generated by a state change on just one edge of $T$ (the central edge), while on $T'$ at least two pendant edges require state changes.
 Thus,  for suitable constants $c,c'$,  for the tree in Fig. 1(a) we have: $p_i \geq \frac{c}{i}$ and $p_i' \leq c'(\frac{L}{i})^2$;
while for the tree in Fig. 1(b) we have $p_i \geq \frac{c}{i}$ and $p_i' \leq c'( \frac{L}{i})( \frac{1}{i})$.
Thus, in either case, provided $i$ is sufficiently large, Inequality (\ref{exq2}) gives:
\begin{equation}
\label{dbound} 
d_i^2 \geq  \left(\frac{\sqrt{c}}{\sqrt{i}} - \frac{\sqrt{c'}L}{i}\right)^2 \geq  \frac{d}{i}(1-o(1)),
\end{equation}
 for a positive constant $d$, and where $o(1)$ denotes a term that converges to $0$ with increasing $i$. Then, by combining  Eqn. (\ref{DHeq}) and Eqn. (\ref{dbound}), we have:\begin{equation}
\label{exq}
2 \geq D_H^2(T_1, T') \geq 2\left(1-\prod_{i=L^4}^{L^{5}} (1-\frac{1}{2}d_i^2)\right) \geq 2\left(1-\prod_{i=L^4}^{L^{5}}(1-\frac{d(1-o(1))}{2i})\right),
\end{equation}
and straightforward asymptotic analysis of the last term reveals that  $D_H^2(T_1, T') \rightarrow  2$ as $L \rightarrow \infty$.

Finally, we invoke an inequality (Theorem 3.2) from \cite{SZE}.  If $M= $MLE (maximum likelihood estimation) then for $A=\{T_1, T'\}$, the probability that MLE correctly reconstructs the generating tree from $A$ is at least $\frac{1}{2}D_H(T_1, T')$ and this converges to $1$ as $L$ grows (with $k$ growing at the rate $L^5$).

\subsection{Breaking up long edges by adding more taxa}

Based on simulation studies and qualitative understanding, it is received wisdom that long branches are untrustworthy due to the long branch being able to `go anywhere'.  Hence biologists seek to break up this branch either with more characters or more taxa (see, for example, \cite{Fels} or \cite{Graybeal}).  One could then reconstruct a phylogenetic tree for this `extended' set of taxa and then ignore all but the few taxa one is interested in.  

However, as we add more taxa, the  number of possible phylogenetic trees grows exponentially, and more data are required to
reconstruct a larger tree correctly (this can easily be seen by a purely counting argument). Thus it is not immediately clear whether this strategy has any formal basis for improving accuracy.  Here, we show that the sequence length requirements   for resolving a four-taxon tree that has one long branch  can be exponentially (or even double-exponentially) greater than those of the large tree in certain ideal situations.

To see this, suppose we have one of the types of trees shown in Fig. 1, with one or  more long pendant edges of length $L$. Suppose one can find a set $S$  of $N$ additional taxa so that each edge in the resulting tree has a branch length that lies between fixed values, say $l$ and $l'$ (with $l\leq l_0\leq l'$).  Reconstructing this larger tree accurately requires just some constant times $\log(N)/l^2$ sites under a two-state symmetric model (see \cite{das}) provided $l'$ lies below a critical transition value,  while reconstructing the four-taxon tree involves a term  ($e^{cL}$) that grows exponentially with the length $L$ of any long edge (by Theorem~\ref{thm1}).

The significance of this result hinges on the following question: how does  $\log(N)$ compare with $e^{cL}$?  If very short
branches are attached at equally spaced intervals along the long pendant branch (or branches), then $N$ grows in a linear relationship with $L$.  In this case, the sequence length required to reconstruct the
four-taxon tree is {\em doubly exponential} in the sequence length required to reconstruct the much larger tree,  as $L$ grows (moreover, this does not require the strong technical result from \cite{das} but a weaker though more
generally applicable result from \cite{logs}). 

However, it would be more realistic to constrain the branch lengths in the tree to be approximately clocklike. In that case, $N$ need only be of order $2^{dL}$ for some constant $d$; $\log(N)$ would then be proportional to $L$ and the sequence length required to reconstruct the
four-taxon tree would be exponential in the sequence length required to reconstruct the much larger tree as $L$ grows.  

In this analysis we are, of course,  assuming the most ideal situation, where the
taxa are distributed as favorably as possible to allow the large tree to be reconstructed; still,  it is interesting to note that this route -- constructing a large tree accurately,  then ignoring the majority of taxa  to consider just the induced phylogenetic relationship between four taxa -- can require much shorter sequences lengths to achieve the same accuracy (and this holds for  statistically consistent tree reconstruction methods, not just for inconsistent methods that can be `misled' by long branches).

\subsection{Extension of Theorem~\ref{thm1} to trees with more taxa}
 
Finally, we discuss what happens to our main results concerning four-taxon trees  if we replace one or more of the four leaves of the tree by subtrees.   Firstly, the lower bound on $k$ given by Theorem~\ref{thm1} still applies if the $l_i$ values refer to the lengths of the central three edges. This is because the sequences at the root of the four subtrees screens off the states of the leaves from the random variable that is the  topology $T$ of the central part of the tree (by the Markov property).  More formally, consider the following two data sets:
\begin{itemize}
\item the  sequences $Z$ at the leaves of the tree;
\item the sequences $Y$ at the roots of the four subtrees;
\end{itemize}
Since  $T \rightarrow Y \rightarrow Z$ is a Markov chain, the
`data processing inequality' (\cite{cov}) ensures that   $I(T; Z) \leq I(T; Y),$ 
where $I$ refers to mutual information.  
In other words, the information that the leaves of the tree tell us about $T$ (the topology of the central part of tree) cannot exceed the information that the ancestral sequences at the roots of those subtrees provide about $T$ (were these known; recall that we only observe sequences at the leaves of the tree).   Thus we obtain a conservative lower bound on the required sequence length with these considerations. 

However, a tighter bound would presumably take into account how much uncertainty there is in the state at the root of one of the four subtrees, given the states we observe at the leaves of that tree.

To simplify the discussion here, consider just the symmetric two-state model of site substitution. In this case, let $p_i$ denote the probability of accurately inferring the root state of a subtree that stands in place of taxon $i$ from the states at the leaves under maximum likelihood (we assume that the topology and branch lengths of the subtree are known).  If $l_i$ is the length of the central branch of $T$ that is incident with the root of this subtree, then the probability of a substitution across the endpoints of this edge is
$p'_i = \frac{1}{2}(1-e^{-2l_i})$.   

This suggests the possibility of approximating the sequence length required to resolve a polytomy in a large tree by replacing each of the four incident subtrees by a single taxon, with a net probability of substitution across branch $i$  being set to $p_i\cdot p'_i$. Thus we have replaced a phylogenetic tree with four subtrees by a four-taxon tree, in which the central edge is of the same length but the
pendant edges have been `lengthened' to allow for the loss of information that the leaves provide concerning the root state of each subtree. A natural candidate for this 
`effective branch length' of branch $i$ would be a value of $l$ for which $p_i\cdot p'_i = \frac{1}{2}(1-e^{-2l})$; this  has the solution:
$l=l_i + \frac{1}{2}\log(1/(1-2p_i)).$   It may be interesting to explore this approach further  since the computation and behavior of the expected root-state reconstruction probability ($p_i$) have been analyzed already by a number of authors (e.g. \cite{eva}, and \cite{ma2}).

\section{Acknowledgments}  We thank the {\em Allan Wilson Centre for Molecular Ecology and Evolution} for supporting this work.

\bibliographystyle{model2-names}
\bibliography{SequenceBounds}

\begin{thebibliography}{19}
\expandafter\ifx\csname natexlab\endcsname\relax\def\natexlab#1{#1}\fi
\expandafter\ifx\csname url\endcsname\relax
  \def\url#1{\texttt{#1}}\fi
\expandafter\ifx\csname urlprefix\endcsname\relax\def\urlprefix{URL }\fi
\providecommand{\eprint}[2][]{\url{#2}}
\providecommand{\bibinfo}[2]{#2}
\ifx\xfnm\relax \def\xfnm[#1]{\unskip,\space#1}\fi
\bibitem[{Aldous and Fill(2010)}]{Ald}
\bibinfo{author}{Aldous, D.}, \bibinfo{author}{Fill, J.}, \bibinfo{year}{2010}.
\newblock \bibinfo{title}{Reversible Markov chains and random walks on graphs}.
\bibitem[{Churchill et~al.(1992)Churchill, von Haeseler and Navidi}]{chu}
\bibinfo{author}{Churchill, G.}, \bibinfo{author}{von Haeseler, A.},
  \bibinfo{author}{Navidi, W.}, \bibinfo{year}{1992}.
\newblock \bibinfo{title}{Sample size for a phylogenetic inference}.
\newblock \bibinfo{journal}{Mol. Biol. Evol.} \bibinfo{volume}{9},
  \bibinfo{pages}{753--769}.
\bibitem[{Cover and Thomas(1991)}]{cov}
\bibinfo{author}{Cover, T.}, \bibinfo{author}{Thomas, J.},
  \bibinfo{year}{1991}.
\newblock \bibinfo{title}{Elements of Information Theory}.
\newblock \bibinfo{publisher}{Wiley, New York}.
\bibitem[{Daskalakis et~al.(2011)Daskalakis, Mossel and Roch}]{das}
\bibinfo{author}{Daskalakis, C.}, \bibinfo{author}{Mossel, E.},
  \bibinfo{author}{Roch, S.}, \bibinfo{year}{2011}.
\newblock \bibinfo{title}{Evolutionary trees and the {I}sing model on the
  {B}ethe lattice: a proof of {S}teel's conjecture}.
\newblock \bibinfo{journal}{Probab. Theor. Relat. Field} \bibinfo{volume}{149},
  \bibinfo{pages}{149--189}.
\bibitem[{Erd\"{o}s et~al.(1999)Erd\"{o}s, Steel, Sz\'{e}kely and
  Warnow}]{logs}
\bibinfo{author}{Erd\"{o}s, P.L.}, \bibinfo{author}{Steel, M.A.},
  \bibinfo{author}{Sz\'{e}kely, L.}, \bibinfo{author}{Warnow, T.},
  \bibinfo{year}{1999}.
\newblock \bibinfo{title}{A few logs suffice to build (almost) all trees (part
  2)}.
\newblock \bibinfo{journal}{Theor. Comput. Sci.} \bibinfo{volume}{221},
  \bibinfo{pages}{77--118}.
\bibitem[{Evans et~al.(2000)Evans, Kenyon, Peres and Schulman}]{eva}
\bibinfo{author}{Evans, W.}, \bibinfo{author}{Kenyon, C.},
  \bibinfo{author}{Peres, Y.}, \bibinfo{author}{Schulman, L.},
  \bibinfo{year}{2000}.
\newblock \bibinfo{title}{Broadcasting on trees and the {I}sing model}.
\newblock \bibinfo{journal}{Ann. Appl. Probab.} \bibinfo{volume}{10},
  \bibinfo{pages}{410--433}.
\bibitem[{Felsenstein(2004)}]{Fels}
\bibinfo{author}{Felsenstein, J.}, \bibinfo{year}{2004}.
\newblock \bibinfo{title}{Inferring Phylogenies}.
\newblock \bibinfo{publisher}{Sinauer Associates, Sunderland, MA}.
\bibitem[{Fischer and Steel(2009)}]{FS}
\bibinfo{author}{Fischer, M.}, \bibinfo{author}{Steel, M.},
  \bibinfo{year}{2009}.
\newblock \bibinfo{title}{Sequence length bounds for resolving a deep
  phylogenetic divergence}.
\newblock \bibinfo{journal}{J. Theor. Biol.} \bibinfo{volume}{256},
  \bibinfo{pages}{247--252}.
\bibitem[{Graybeal(1998)}]{Graybeal}
\bibinfo{author}{Graybeal, A.}, \bibinfo{year}{1998}.
\newblock \bibinfo{title}{Is it better to add taxa or characters to a difficult
  phylogenetic problem?}
\newblock \bibinfo{journal}{Syst. Biol.} \bibinfo{volume}{47},
  \bibinfo{pages}{9--17}.
\bibitem[{Lecointre et~al.(1994)Lecointre, Philippe, Van~Le and
  Le~Guyader}]{lec}
\bibinfo{author}{Lecointre, G.}, \bibinfo{author}{Philippe, H.},
  \bibinfo{author}{Van~Le, H.}, \bibinfo{author}{Le~Guyader, H.},
  \bibinfo{year}{1994}.
\newblock \bibinfo{title}{How many nucleotides are required to resolve a
  phylogenetic problem? {T}he use of a new statistical method applicable to
  available sequences}.
\newblock \bibinfo{journal}{Mol. Phyl. Evol.} \bibinfo{volume}{3},
  \bibinfo{pages}{292--309}.
\bibitem[{Ma and Zhang(2011)}]{ma2}
\bibinfo{author}{Ma, B.}, \bibinfo{author}{Zhang, L.}, \bibinfo{year}{2011}.
\newblock \bibinfo{title}{Efficient estimation of the accuracy of the maximum
  likelihood method for ancestral state reconstruction}.
\newblock \bibinfo{journal}{J. Combin. Optimization} \bibinfo{volume}{21},
  \bibinfo{pages}{409--422}.
\bibitem[{McDiarmid(1989)}]{mcd}
\bibinfo{author}{McDiarmid, C.}, \bibinfo{year}{1989}.
\newblock \bibinfo{title}{On the method of bounded difference}, in:
  \bibinfo{booktitle}{Surveys in Combinatorics}. \bibinfo{publisher}{Cambridge
  University Press, Cambridge}, pp. \bibinfo{pages}{148--188}.
\bibitem[{Mossel and Steel(2005)}]{mos}
\bibinfo{author}{Mossel, E.}, \bibinfo{author}{Steel, M.},
  \bibinfo{year}{2005}.
\newblock \bibinfo{title}{How much can evolved characters tell us about the
  tree that generated them?}, in: \bibinfo{editor}{Gascuel, O.} (Ed.),
  \bibinfo{booktitle}{Mathematics of Evolution and Phylogeny}.
  \bibinfo{publisher}{Oxford University Press, Oxford}, pp.
  \bibinfo{pages}{384--412}.
\bibitem[{Rozanov(1969)}]{Roz}
\bibinfo{author}{Rozanov, Y.}, \bibinfo{year}{1969}.
\newblock \bibinfo{title}{Probability Theory: A Concise Course}.
\newblock \bibinfo{publisher}{Dover Publications}.
\bibitem[{Saitou and Nei(1986)}]{sai}
\bibinfo{author}{Saitou, N.}, \bibinfo{author}{Nei, M.}, \bibinfo{year}{1986}.
\newblock \bibinfo{title}{The number of nucleotides required to determine the
  branching order of three species, with special reference to the
  human-chimpanzee-gorilla divergence}.
\newblock \bibinfo{journal}{J. Mol. Evol.} \bibinfo{volume}{24},
  \bibinfo{pages}{189--204}.
\bibitem[{Steel and Sz\'{e}kely(2002)}]{SZE}
\bibinfo{author}{Steel, M.}, \bibinfo{author}{Sz\'{e}kely, L.},
  \bibinfo{year}{2002}.
\newblock \bibinfo{title}{Inverting random functions {II}: explicit bounds for
  discrete maximum likelihood estimation, with applications}.
\newblock \bibinfo{journal}{SIAM. J. Discr. Math.} \bibinfo{volume}{15},
  \bibinfo{pages}{562--575}.
\bibitem[{Townsend(2007)}]{tow07}
\bibinfo{author}{Townsend, J.}, \bibinfo{year}{2007}.
\newblock \bibinfo{title}{Profiling phylogenetic informativeness}.
\newblock \bibinfo{journal}{Syst. Biol.} \bibinfo{volume}{56},
  \bibinfo{pages}{222--231}.
\bibitem[{Townsend and Leuenberger(2011)}]{tow11}
\bibinfo{author}{Townsend, J.}, \bibinfo{author}{Leuenberger, C.},
  \bibinfo{year}{2011}.
\newblock \bibinfo{title}{Taxon sampling and the optimal rates of evolution for
  phylogenetic inference}.
\newblock \bibinfo{journal}{Syst. Biol.} \bibinfo{volume}{60},
  \bibinfo{pages}{358--365}.
\bibitem[{Townsend et~al.(2012)Townsend, Su and Tekle}]{tow12}
\bibinfo{author}{Townsend, J.}, \bibinfo{author}{Su, Z.},
  \bibinfo{author}{Tekle, Y.}, \bibinfo{year}{2012}.
\newblock \bibinfo{title}{Phylogenetic signal and noise: predicting the power
  of a data set to resolve phylogeny (in press)}.
\newblock \bibinfo{journal}{Syst. Biol.} .

\end{thebibliography}

\end{document}